\documentclass[pre,aps,preprint,showpacs]{revtex4}
\usepackage{revsymb,latexsym,amssymb,amsmath,comment}
\usepackage{graphicx,psfrag,subfigure,stmaryrd}

\begin{document}
\author{J. M. Wesselinowa }
\affiliation{University of Sofia, Department of Physics\\ Blvd. J. Bouchier 5, 
1164 Sofia, Bulgaria}
\email{julia@phys.uni-sofia.bg}
\author{S.Trimper, and K. Zabrocki}
\affiliation{Fachbereich Physik,
Martin-Luther-Universit\"at, D-06099 Halle Germany}
\email{trimper@physik.uni-halle.de}
\title{Impact of layer defects in ferroelectric thin films}
\date{\today }
\begin{abstract}
Based on a modified Ising model in a transverse field we demonstrate that defect layers in ferroelectric 
thin films, such as layers with impurities, vacancies or dislocations, are able to induce a strong 
increase or decrease of the polarization depending on the variation of the exchange interaction within the 
defect layers. A Green's function technique enables us to calculate the polarization, the excitation energy 
and the critical temperature of the material with structural defects. Numerically we find the polarization 
as function of temperature, film thickness and the interaction strengths between the layers. The theoretical 
results are in reasonable accordance to experimental datas of different ferroelectric thin films. 


\end{abstract}

\maketitle

\section{Introduction}
Defects in crystals influence significantly the physical properties of almost all materials. 
In the last decades defect engineering has been developed as a part of modern semiconductor 
technology. There is also an increasing interest in studying defects and their related strain 
fields in ferroelectrics (FE), in particular in low dimensional systems \cite{1}. A broad variety of 
defects have been implemented and analyzed mainly in ABO$_3$-type FE, such as BaTiO$_3$ and Pb(Zr,Ti)O$_3$ (PZT) 
\cite{1,2,3,4}, bismuth layer-structured FE, such as Bi$_4$Ti$_3$O$_{12}$ (BTO) \cite{5,6,7,8} as well as  
SrBi$_2$Ta$_2$O$_9$ (SBT) \cite{9,10,11}. To control the polarization and piezoelectric properties,
PZT is often modified by various elements of lower valency (K, Mn, Fe) and higher talent 
(La, Nb, Ta) cations. The modified PZT are generally classified into the two categories named as  
"hard" and "soft" ones \cite{12}. The substitution of lower-valent cations, compared to the 
constituent ions, leads to oxide vacancies which induces a soft FE behavior, whereas  
higher-valent substitution gives rise to cation vacancies offering a hard FE behavior. 
Contrary to hard FE, the soft FE are characterized by a lower coercitive field $E_c$, lower remanent 
polarization $\sigma _r$ and higher hysteresis losses. The different FE properties are generally originated by 
the occurrence of defects. In materials with confined dimensions, the influence of defects can be even more 
pronounced due to the enhanced relative volume of the "defective" regions \cite{4}. 

FE thin films have been extensively studied due to their ability for applications in
nonvolatile FE random access memories (FeRAMs) \cite{13}. For this application the thin film 
ferroelectrics should reveal large remanent polarizations, low coercitive fields, 
and fatigue-free properties. PZT and SBT have been widely investigated as appropriate material 
in Fe-RAMs. Otherwise the poor fatigue characteristic and the small remanent polarization, are 
viewed as the major barrier for their applications. 

BTO is a promising alternative to PZT, because of its large spontaneous polarization 
along the $a$-axis. Recently, it was reported that the substitution of small amounts 
of impurities, such as Sm and Nd for Bi and V, W and Nb for Ti in the 
pseudoperowskite [(Bi$_2$Ti$_3$O$_{10})^{2-}$] layers of BTO, is effective for lowering the 
leakage current and enhancing the remanent polarization \cite{6,7,8}. 
Noguchi et al. \cite{9} reported the control of FE properties in Sr-site-modified SBT bulk
ceramics, in which the Sr-site was partially substituted by La atoms, showed a lower 
coercitive field as the original material; but, the remanent polarization was also 
decreased in this material. On the other hand, several reports have revealed that the 
$\sigma _r$ value is enhanced in Sr-deficient and Bi-excess SBT films because of the 
substitution of excess Bi atoms in the Sr site \cite{14,15}. However, $E_c$ also increases in
these films. Since the ionic radius of the La ion is almost the same as that of the 
Bi ion, the electronegativity is also important for controlling the FE properties in
the site-engineering technique. In fact, the electronegativity of a Bi ion is much 
higher than that of a La ion. A similar effect was obtained in Sr-deficient and Pr-substituted SBT 
films \cite{10}. 

PZT films are considered to most promising candidates for applications in microelectromechanical 
systems (MEMS) since they have large piezoelectric coefficients and electromechanical coupling
coefficients. As known from bulk ceramics, the piezoelectric and dielectric properties vary 
substantially with the composition (Zr to Ti ratio). While the bulk PZT ceramics have been well 
investigated, there is still a lack of understanding the piezoelectric and dielectric 
properties of PZT thin films, particularly as a function of composition. With increasing 
Ti-content, larger remanent polarization and higher coercive voltage are observed \cite{1}. 
Impact of misfit dislocations on the polarization instability of epitaxial nanostructured 
FE perovskites (PZT nanoislands) is investigated by Chu et al. \cite{4}. Their results suggest 
that misfit engineering is indispensable for obtaining nanostructured FE's with stable 
polarization. The properties of BaTiO$_3$ are modified by substituting of tetravalent ions such as 
Zr$^{4+}$ for Ti sites. This increases the dielectric constant and broadens the sharp 
temperature dependence of the dielectric constant around the Curie temperature. 
With the addition of strontium to PbTiO$_3$, the phase-transition temperature decreases with 
the increasing Sr concentration.

Influence of defects on FE phase transitions is one of key-like topics of the modern studies 
of FE. Constant attention to this topic is connected with its role in the real FE problem 
solution from the point of views of the fundamental science as well as of the above mentioned 
applications. It is well known that localized spin excitations can arise in FE and magnetic 
systems with broken spatial translational symmetry, i.e. due to the presence of boundary 
surfaces, interfaces, impurities and other defects in the crystal. These localized modes 
appearing in the above mentioned structures are experimentally observed using the neutron 
and Raman scattering \cite{16,17}, or, more recently, far infrared absorption \cite{18}.
The Green's function theories have been extensively employed to describe these modes in 
semi-infinite transverse Ising models \cite{19} and FE thin films \cite{20}. In addition to bulk spin 
waves there may be localized modes associated with the impurity layer \cite{19} and with the 
surface \cite{20}. A transfer matrix formalism is applied to study a semi-infinite Heisenberg 
ferromagnetic superlattice with surface cell and impurity cell at arbitrary distance from 
the surface \cite{21}. The occurrence of localized modes associated with the impurity cell is 
demonstrated sufficiently. In the dielectric continuum approximation, Zhang et al. \cite{22} 
investigated the effect of different kinds of coupling to the localized interface optical-phonon modes in two 
coupled semi-infinite $N$-constituent superlattices with a structural defect layer.
The influence of molecular impurity ions on FE phase transitions is studied by Vikhin and 
Maksimova using the phenomenological Landau theory \cite{23}. It is shown that molecular impurity
ions with charge transfer-local vibration degrees of freedom which bi-linearly interact with 
FE order parameter can induce rather strong increasing of the critical temperature of FE 
phase transition at reasonable concentrations of molecular impurities. It could be topical, 
for instance, for the KDP:MnO$_4$ case. Using a combination of first-principles and 
effective-Hamiltonian approaches, Dieguez et al. \cite{24} obtained a phase diagram in temperature 
and misfit strain in epitaxial BaTiO$_3$ that is qualitatively different from that reported 
by Pertsev et al. \cite{25}. In particular, in \cite{24} is found a region of "$r$ phase" at low 
temperature where Pertsev et al. have reported an "$ac$ phase". 

The most of the published papers consider the defects influence on the spin excitations. 
A first attempt to find out the influence of defects on the polarization of FE thin films 
are elucidated by Alpay et al. \cite{26,27}. A thermodynamic analysis has been carried out to 
investigate the role of dislocations in FE materials. Due to the coupling of the stress 
field of the dislocation and the polarization, there is expected a drastic variation of the 
polarization near to the dislocation.
The compressive regions enhance the polarization and increase Curie temperature whereas
tensile stresses decrease polarization with a commensurate drop in the Curie temperature 
\cite{26,27}. But actually the dependence of polarization and Curie temperature on defects in 
FE thin films is not so intensively studied theoretically. Therefore, it is the aim of the 
present paper to investigate the polarization of FE thin films with defect layers in more 
detail. 
Based on the Ising model in a transverse field and using the Green's function theory we 
calculate the spin-wave energies, the polarization and the phase transition temperature 
for a FE thin film with different structural defect layers. The numerical results are compared with 
those of bulk materials and with thin films without defect layers.

\section{The Model and the Matrix Green's Function}
Let us consider a three dimensional ferroelectric system on a simple cubic lattice composed 
of $N$ layers in z-direction. The layers are numbered by $n=1,\ldots,N$, where the layers $n=1$ and 
$n=N$ represent the two surfaces of the system. The bulk is established by the remaining 
$(N-2)$ layers. The specific surface effects are included by additional coupling parameters 
between bulk and surface layers. In particular, we start with the Hamiltonian of the Ising 
model in a transverse field which includes both, bulk and surface properties.
\begin{equation}
H=-\frac{1}{2}\sum_{ij}J_{ij}S^z_iS^z_j-\Omega_b\sum_{i\epsilon b}S^x_i-
\Omega_s\sum_{i\epsilon s}S^x_i,
\label{mo1}
\end{equation}
where $S^x$ and $S^z$ are components of spin-$\frac{1}{2}$ operators, and 
$\Omega_b$ and $\Omega_s$ represent transverse fields in the bulk and surface 
layers, and the sums are over the internal and surface lattice points, 
respectively. $J_{ij}$ is an exchange interaction between spins at 
nearest-neighbor sites $i$ and $j$, and $J_{ij}=J_s$ between spins on the 
surface layer, otherwise it is $J_b$. 
We assume that one or more of the layers can be defect, since 
$J_d$ and $\Omega_d$ denote the exchange 
interaction and the transverse field of the defect layer. In the ordered phase 
we have the mean values $<S^x>\neq0$ and $<S^z>\neq0$, and it is appropriate to 
choose a new coordinate system rotating the original one used in (1) by the angle 
$\theta$ in the $xy$ plane \cite{20}. The rotation angle $\theta$ is determined by the requirement
$<S^{x'}>=0$ in the new coordinate system. 

The retarded Green's function to be calculated is defined as
\begin{equation}
G_{ij}(t)=\ll{S^+_i(t);S^-_j(0)}\gg,
\label{gf1}
\end{equation}
where $S^+$ and $S^-$ are the spin-$\frac{1}{2}$ operators in the rotated system.
On introducing the two-dimensional Fourier transform $G_{n_in_j}({\bf k}_
\parallel,\omega)$, one has the following form:
\begin{equation}
\ll{S^+_i;S^-_j}\gg_{\omega}=\frac{\sigma}{N'}\sum_{{\bf k}
_{\parallel}}\exp(i{\bf k}
_{\parallel}({\bf r}_i-{\bf r}_j))G_{n_in_j}({\bf k}_{\parallel}, \omega),
\label{gf2}
\end{equation}
where $N'$ is the number of sites in any of the lattice planes, ${\bf r}_i$ 
and $n_i$ represent the position vectors of site i and the layer index, 
respectively, ${\bf k}_{\parallel}=(k_x,k_y)$ is a two-dimensional wave vector
parallel to the surface. The summation is taken over the Brillouin zone.

As a result the equation of motion for the Green's function in Eq.~(\ref{gf2}) of the ferroelectric
thin film for $T \leq T_c$ has the following matrix form:
\begin{equation}
{\bf H}(\omega){\bf G}({\bf k}_{\parallel},\omega)={\bf R},
\label{gf3}
\end{equation}
where ${\bf H}$ can be expressed as:
\begin{displaymath}
\mathbf{H} =
\left( \begin{array}{ccccccc}
\omega-V_1 &   k_1      &    0       &    0   &    0   &    0   & \ldots \\
    k_2    & \omega-V_2 &   k_2      &    0   &    0   &    0   & \ldots \\
   0     &   k_3      & \omega-V_3 &   k_3  &    0   &    0   & \ldots \\
      \vdots  & \vdots & \vdots & \vdots & \vdots & \vdots & \ddots\\
0 & 0 & 0 & 0 & 0 & k_N & \omega-V_N \\ 
\end{array} \right)
\end{displaymath}
with
\begin{eqnarray*}
k_n&=& J_b \sigma_n \sin^2{\theta_n}, \qquad n=1,...,N, \\
V_1&=&2\Omega_s \sin{\theta}_1 + \frac{1}{2} \sigma_1 J_s \cos^2{\theta}_1-
\frac{\sigma_1 J_s}{4} \sin^2{\theta}_1 \gamma({\bf k_{\parallel}})+
J_b\sigma_2 \cos^2{\theta}_2, \\ 
V_2&=&2\Omega_b \sin{\theta}_2 + \frac{1}{2} \sigma_2 J_b \cos^2{\theta}_2-
\frac{\sigma_2 J_b}{4} \sin^2{\theta}_b \gamma({\bf k_{\parallel}})+J_s \sigma_1 
\cos^2{\theta}_+J_b\sigma_3 \cos^2{\theta}_3, \\
V_n&=&2\Omega_n \sin{\theta}_n + \frac{1}{2} \sigma_n J_b \cos^2{\theta}_n-\frac{\sigma_n J_n}
{4} \sin^2{\theta}_n \gamma({\bf k_{\parallel}})+J_{n-1}\sigma_{n-1} \cos^2{\theta}_{n-1}
\\ &+&J_{n+1}\sigma_{n+1} \cos^2{\theta}_{n+1}, \\
V_N&=&2\Omega_s \sin{\theta}_N + \frac{1}{2} \sigma_N J_s \cos^2{\theta}_N-
\frac{\sigma_N J_s}{4} \sin^2{\theta}_N \gamma({\bf k_{\parallel}})+J_b\sigma_{N-1} 
\cos^2{\theta}_{N-1},\\ 
\gamma({\bf k}_{\parallel})&=&\frac{1}{2}(\cos(k_x a)+\cos(k_y a)).
\end{eqnarray*}
Here we have introduced the notations $J_1 \equiv J_N = J_s$, $J_n=J_b$ for $n=2,3,4,...,N-1$,  
$\Omega_1=\Omega_N=\Omega_s$, $\Omega_n=\Omega_b$ (n=2,3,4,...,N-1), $J_0=J_{N+1}=0$. The quantity $\sigma(T)$ 
is the relative polarization in the direction of the mean field and is equal to 
$2\langle S^{z'} \rangle$. For the rotation angle $\theta$ we have the following
two solution in the generalized Hartree-Fock approximation:
\begin{eqnarray*}
&1.& \cos{\theta}=0, \quad i.e.\quad  \theta=\frac{\pi}{2}, \quad\quad if \quad 
T \geq T_c;\nonumber\\
&2.& \sin{\theta}=\frac{4\Omega}{\sigma J}=\frac{\sigma_c}{\sigma}, 
\quad\quad\quad\quad\quad if \quad T\leq T_c.
\end{eqnarray*}

In order to obtain the solutions of the matrix equation (\ref{gf3}), we introduce the 
two-dimensional column matrices, ${\bf G}_m$ and ${\bf R}_m$, where the elements 
are given by $({\bf G}_{n})_m = G_{mn}$ and 
$({\bf R}_{n})_m = \sigma _n \delta _{mn}$, so that Eq.~(\ref{gf3}) yields
\begin{equation}
{\bf H}(\omega ) {\bf G}_n ={\bf R}_n. 
\label{m}
\end{equation}
From Eq.~(\ref{m}), $G_{nn}(\omega )$ is obtained as:
\begin{equation}
G_{nn}(\omega ) = \frac{\mid H_{nn}(\omega ) \mid}{\mid H(\omega )\mid}.
\label{m1}
\end{equation} 
The quantity $ \mid H_{nn}(\omega ) \mid $ is the determinant made by replacing the $n$-th 
column of the determinant $\mid H(\omega )\mid $ by $R_n$. The poles $\omega _n$ of the 
Green`s function $G_{nn} (\omega )$ can be calculated by solving 
$\mid H(\omega )\mid = 0$. 

The relative polarization of the of the $n$-th layer is given by
\begin{equation}
\sigma_n=\Big(\frac{\sigma_nJ_n}{2N}\sum_{{\bf k}_{\parallel}}\frac{1-
0.5\sin^2{\theta_n} \gamma({\bf k}_{\parallel})}{\omega_n}
\coth{\frac{\omega_n}{2T}}\Big)^{-1}.
\label{m2}
\end{equation}
The last equation has to be calculated numerically. Due to the assumption of
symmetrical surfaces, there are $\frac{1}{2}$ N layer polarizations, which have 
to be solved self-consistently. In order to obtain the dependence of the Curie 
temperature $T_C$ on the film thickness $N$, we let all $\sigma$'s be zero in 
Eq.~(\ref{m2}) and solve the expressions self-consistently.

\section{Numerical Results and Discussion} 
\begin{figure} [!ht]
\centering
\psfrag{si}[][][1.3]{${\sigma}$}
\psfrag{Temp}[][][0.75]{$T [K]$}
\includegraphics[scale=0.5]{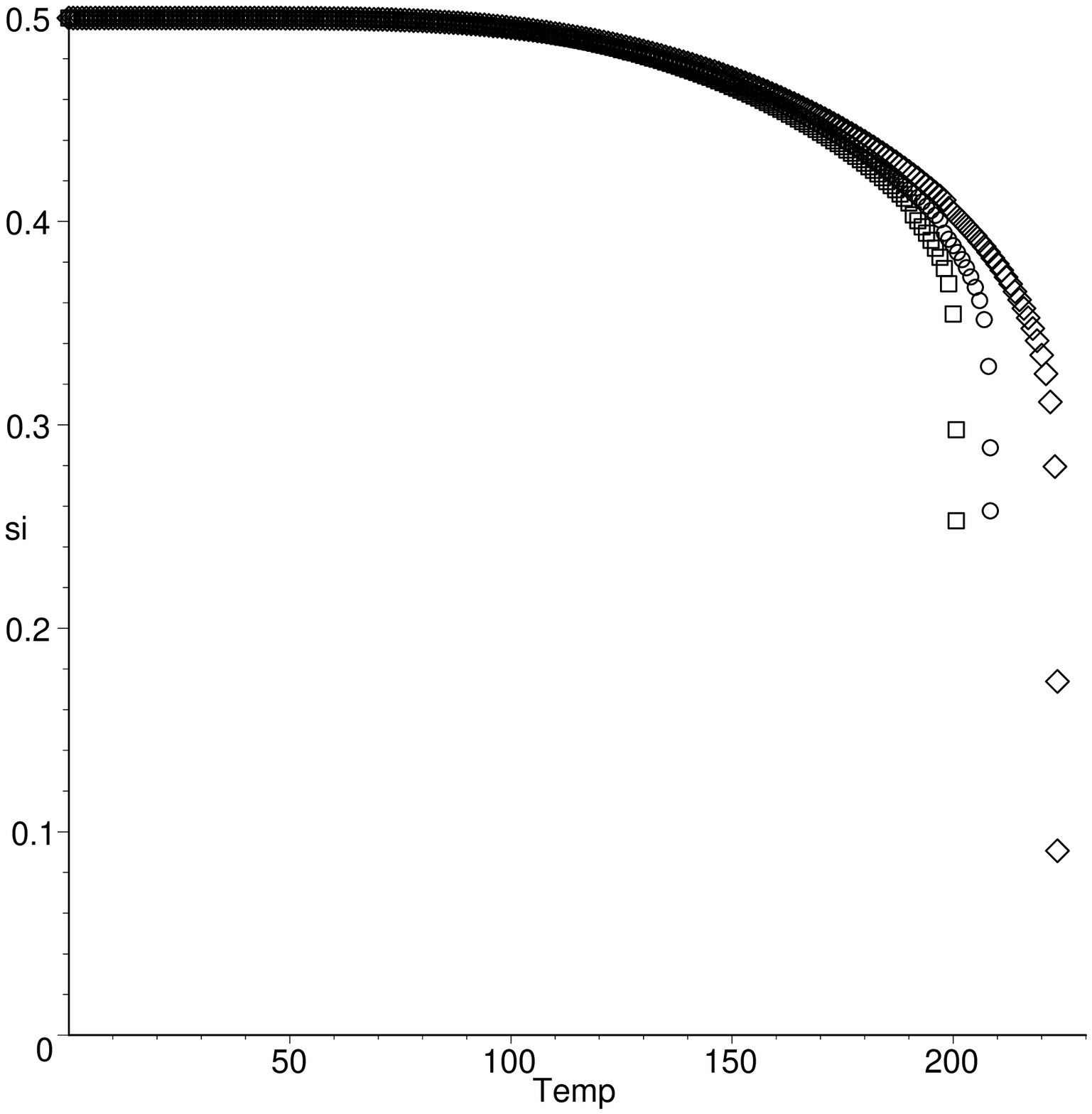}
\caption{Temperature dependence of the polarization ${\sigma}$ for a FE thin film with $J_b = 495 K$, 
$\Omega_b= 20 K$, $J_s=900 K$, $\Omega_s=\Omega_b$, $N$=9 and different $J_d$-values: 
(1) $J_d$=300 K, $T_c =201 K$ ($\boxempty$), (2) $495 K$,\, $T_c =208 K$ $(\circ )$, (3) $1000 K$,\, 
$T_c =224 K$ $(\diamond )$.}
\label{Fig.1}
\end{figure}

\begin{figure} [!ht]
\centering
\psfrag{om}[][][1.1]{$\omega $}
\psfrag{Temp}[][][0.85]{$T [K]$}
\includegraphics[scale=0.5]{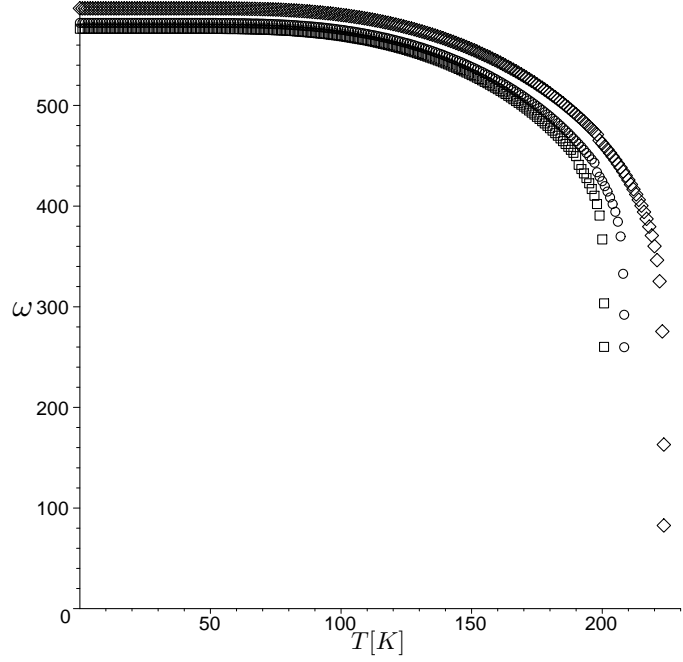}
\caption{Temperature dependence of the spin-wave energy ${\omega} [cm^{-1}]$ for a FE thin film with 
$J_b = 495 K$, $\Omega_b= 20 K$, $J_s=900 K$, $\Omega_s=\Omega_b$, $N$=9 and different 
$J_d$\,-values: (1) $J_d=300 K $ ($\boxempty$), (2) $495 K (\circ )$, (3) $1000 K (\diamond )$.}
\label{Fig.2}
\end{figure}

In this section we present the numerical results based on our theoretical calculations  
taking the typical the bulk parameter $J_b = 495 K$ and $\Omega_b= 20 K$. We have calculated 
the temperature dependence of the polarization defined by 
$$
\sigma \equiv <S^z> = \frac{1}{N} \sum_n \sigma_n(T)\cos {\theta_n}\,,
$$
and the spin-wave frequencies of thin films for different values of the exchange interaction constants. 
One has to solve self-consistently the $\frac{1}{2}$ N coupled Eqs.~(\ref{m2}) to obtain the layer 
polarization from the surface to half of the film (the other half is symmetric due to the assumption of identical surfaces). The numerical results expose some interesting and novel characteristics in the polarization 
and the spin-wave energies in comparison to the case of FE thin films without defects. To characterize 
the complete ferroelectric system both quantities, the polarization $\sigma $ and the spin-wave 
energy $\omega$ , respectively are averaged over the $N$ layers. The results for film thickness $N=9$ 
and different exchange interaction parameters in the 
defect layers $J_d$ are presented in Fig.~\ref{Fig.1} and Fig.~\ref{Fig.2}. Let us firstly consider 
the case that the middle layer is defect layer. For example, it can be originated by localized vacancies or 
impurities with smaller radii and larger distances between them in comparison to the host material. 
It is reasonable to assume that the exchange interaction $J_d$ is smaller than the value of the bulk interaction 
$J_b=495 K$ and has the value $J_5=J_d=300 K$ (compare Figs.~\ref{Fig.1}, \ref{Fig.2} curve 1). The polarization (respectively the spin-wave energy) is smaller than 
the case without defects, $J_5=J_b$ (see Figs.~\ref{Fig.1}, \ref{Fig.2} curve 2). The polarization decreases with 
increasing temperature to vanish at the critical temperature $T_C$ of the thin film.
The critical temperature decreases due to the smaller $J_d$ value. Pontes et al. \cite{3}  
carried out dielectric and Raman spectroscopy studies and obtained that with addition of 
Sr to PbTiO$_3$, the phase-transition temperature decreases with the increasing Sr 
concentration. 

For the case where $J_5=J_d=1000 K$ (Figs.~\ref{Fig.1}, \ref{Fig.2}, curve 3), i.e. $J_d$ is larger than the 
value of the bulk interaction constant $J_b$, (for example when the impurities have a 
larger radius compared with the constituent ions) the polarization (respectively the 
spin-wave energy) is larger than the case without defects, $J_5=J_b$. $T_C$ of the film is 
enhanced in comparison to the bulk value without defects due to the presence of larger
$J_d$ values. This is the opposite behavior compared to the case of
$J_d=300 K$, $J_d<J_b$. Of course, in the case of both strong and small $J_d$ values, 
thin film polarizations tend to the 3D one when $n$ increases, since the excitation 
spectrum varies from 2D to 3D behavior with thickness. The second case, where $J_d>J_b$ 
could explain the experimentally obtained increase in polarization by the substitution of 
impurities, such as Nd, V, W and Nb in the layers of BTO films \cite{6,7,8} 
or by increasing Ca contents in SBT thin films \cite{11}.  

From Figs.~\ref{Fig.1}, \ref{Fig.2} one can observe that the polarization, the spin-wave energies and the 
critical temperature of the FE phase transition are increased or decreased due to different 
exchange interactions in the defect layers. Our results are in qualitatively agreement with 
the experimental data of Noguchi et al. \cite{15}. They have studied SBT films 
($T_C=295^0C$) by the substitution of rare earth cations of La, Ce, Pr, Nd and Sm as well as 
Bi at the A site (Sr site) with Sr vacancies and have shown, that La modification induces 
soft behavior (lower $E_c$ and lower $\sigma _r$), while a large amount of Nd and Sm substitution 
results in a very high $E_c$ (hard), as a result of defect engineering of both Sr and oxide 
vacancies. For not only SBT, but also other bismuth layer-structured FE (BLSF's) $T_C$ is 
strongly influenced by $r_i$ of A-site cations, and BLSF's with smaller A-site cations 
(Ca$^{2+}$) tend to show a higher $T_C$ (420$^0$C). The same amount of larger Ba$^{2+}$ 
brings about a relaxation of FE distortions and resulted in a decrease in $T_C$ to 120$^0$C. 
The substitution of La led to a marked decrease in $T_C$ to 180$^0$C ($x$=0.5),
because the induced A-site vacancies weaken the coupling between neighboring 
BO$_6$ octahedral \cite{5,15}). This result corresponds in our calculations to the case of 
smaller values of the interaction constant in the defect layer $J_d<J_b$. For 
La-modified PbTiO$_3$, $T_C$ decreased significantly, too, with an increase in La content 
\cite{28}. For Bi-SBT $T_C$ rose strongly to 405$^0$C ($x$=0.2) \cite{15}. 
The increase in $T_C$ by Bi substitution is an opposite tendency to that in the case for La-SBT. The dominant role 
plays here the bonding characteristics with oxide ions. The influence of the orbital 
hybridization on $T_C$ is very large, and Bi substitution resulted in a higher $T_C$ \cite{15}.
This experimental result can be described qualitatively good using in our model exchange 
interaction parameters for the defect layer $J_d>J_b$.

\begin{figure} [!ht]
\centering
\psfrag{si}[][][1.3]{${\sigma}$}
\psfrag{Temp}[][][0.85]{$T [K]$}
\includegraphics[scale=0.5]{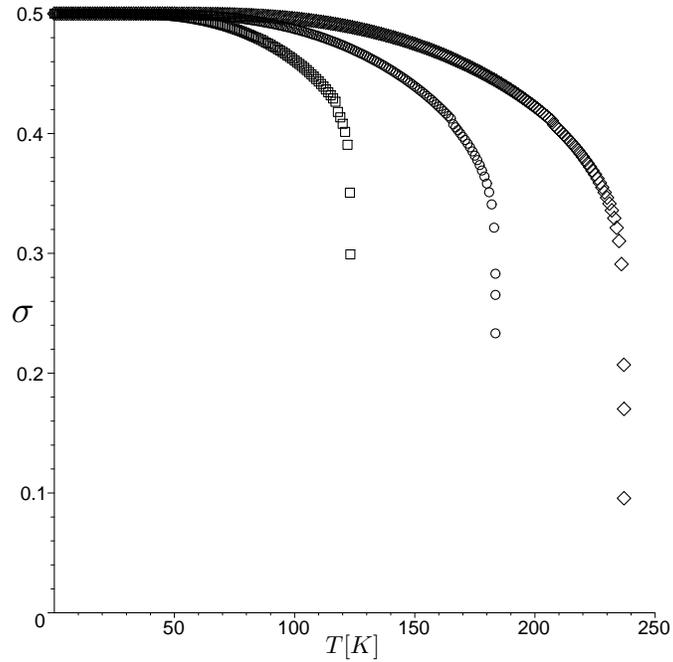}
\caption{Temperature dependence of the polarization $\sigma$ for $J_b = 495 K$, $\Omega_b= 20 K$,
$J_d=1000 K$, $\Omega_s=\Omega_b$, $N$=7 and different $J_s$-values: (1) $J_s =200 K,\, T_c =123 K$ ($\boxempty$), 
(2) $500,\, T_c =184 K$ ($\circ$), 
(3) $1000 K,\, T_c =237 K$ ($\diamond$).}
\label{Fig.3}
\end{figure}

\begin{figure} [!ht]
\centering
\psfrag{om}[][][1.1]{$\omega$}
\psfrag{Temp}[][][0.85]{$T [K]$}
\includegraphics[scale=0.5]{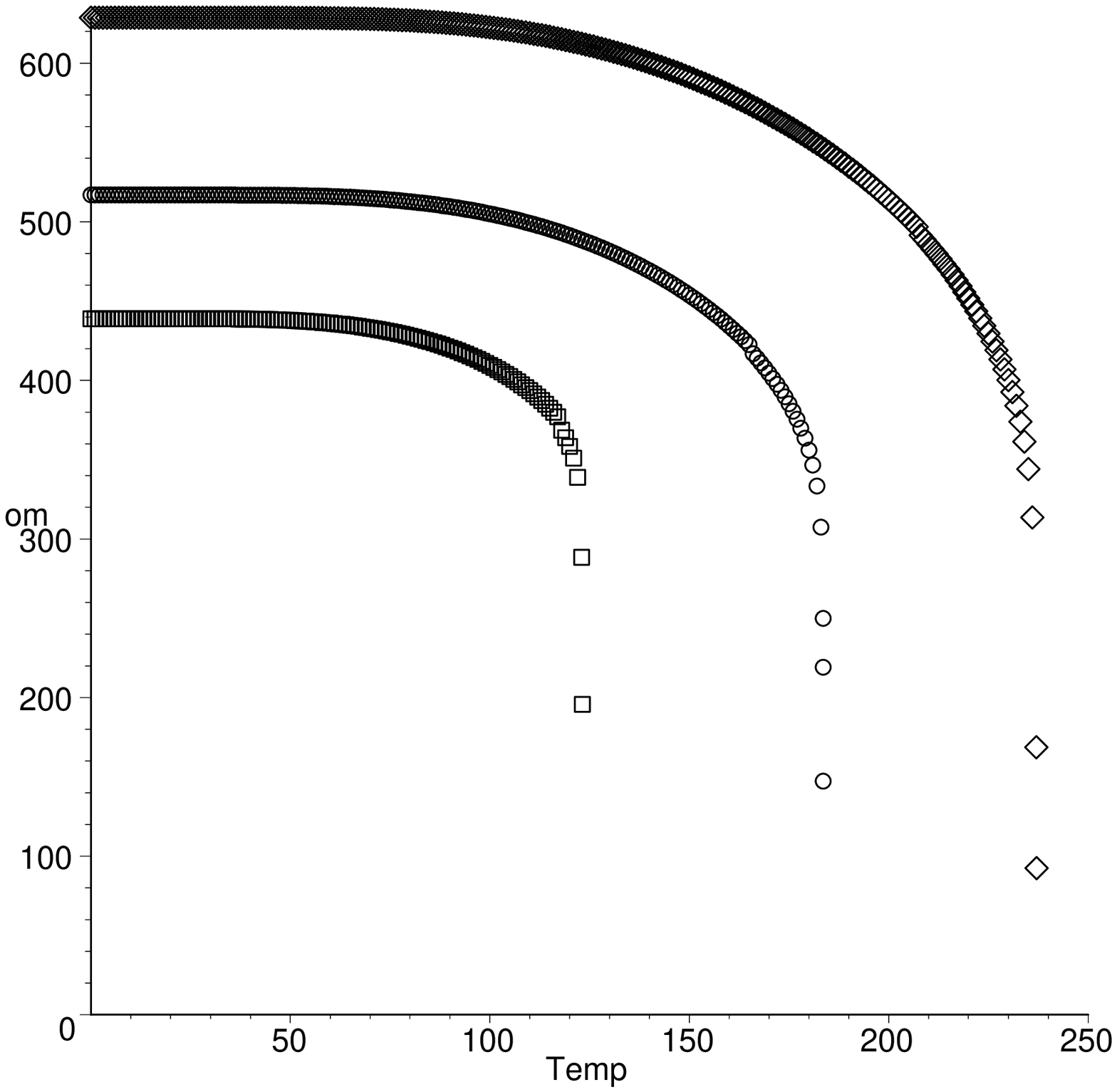}
\caption{Temperature dependence of the spin-wave energy ${\omega} [cm^{-1}]$ for $J_b = 495 K$, 
$\Omega_b= 20 K$, $J_d=1000 K$, $\Omega_s=\Omega_b$, $N$=7 and different $J_s$-values: (1) $J_s =200 K$ ($\boxempty$), (2) $ 500 K (\circ)$, (3) $1000 K $ ($\diamond$).}
\label{Fig.4}
\end{figure}

In Figs.~\ref{Fig.3}, \ref{Fig.4} is shown the influence of the surface exchange interaction constant $J_s$
on the polarization and spin-wave energy by $J_4=J_d$ = const for a thin film with 
$N$ = 7 layers. It can be seen that the two quantities are enhanced with increasing of $J_s$. 
The critical temperature $T_C$ increases, too. 

\begin{figure} [!ht]
\centering
\psfrag{si}[][][1.3]{${\sigma}$}
\psfrag{Temp}[][][0.75]{$T [K]$}
\includegraphics[scale=0.5]{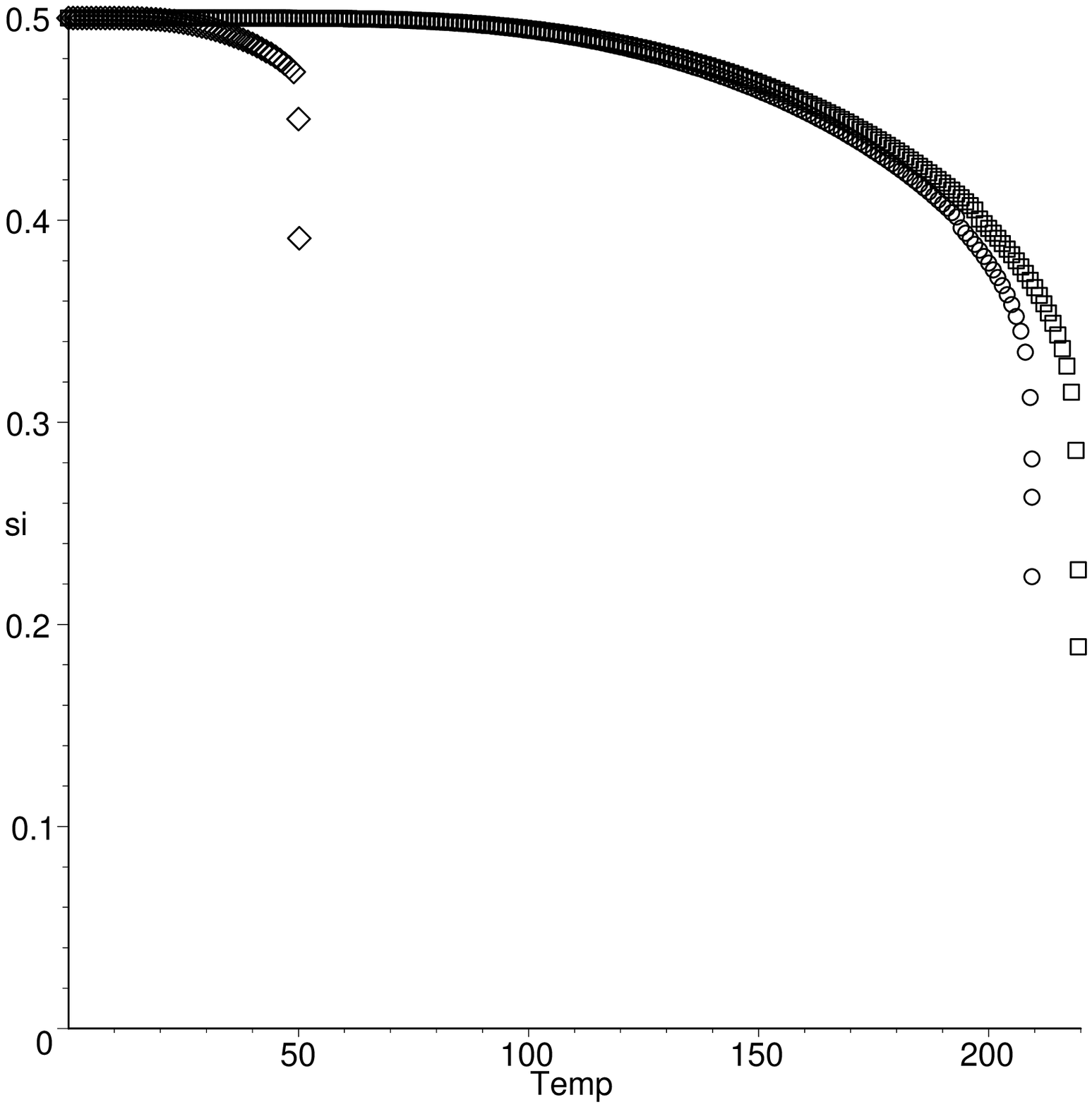}
\caption{Temperature dependence of the polarization ${\sigma}$ for $J_b = 495 K$, $\Omega_b= 20 K$,
$J_s=900 K$, $\Omega_s=\Omega_b$, $N$=7 and different defect layers: (1) $J_4=J_b=495 K,\, T_c =220 K (\boxempty )$, 
(2) $J_4=J_d=200 K,\, T_c = 209 K (\circ )$, (3) $J_3=J_4=J_5=J_d = 200 K,\, T_c =50 K (\diamond )$.}
\label{Fig.5}
\end{figure}

\begin{figure} [!ht]
\centering
\psfrag{si}[][][1.3]{${\sigma}$}
\psfrag{Temp}[][][0.85]{$T [K]$}
\includegraphics[scale=0.5]{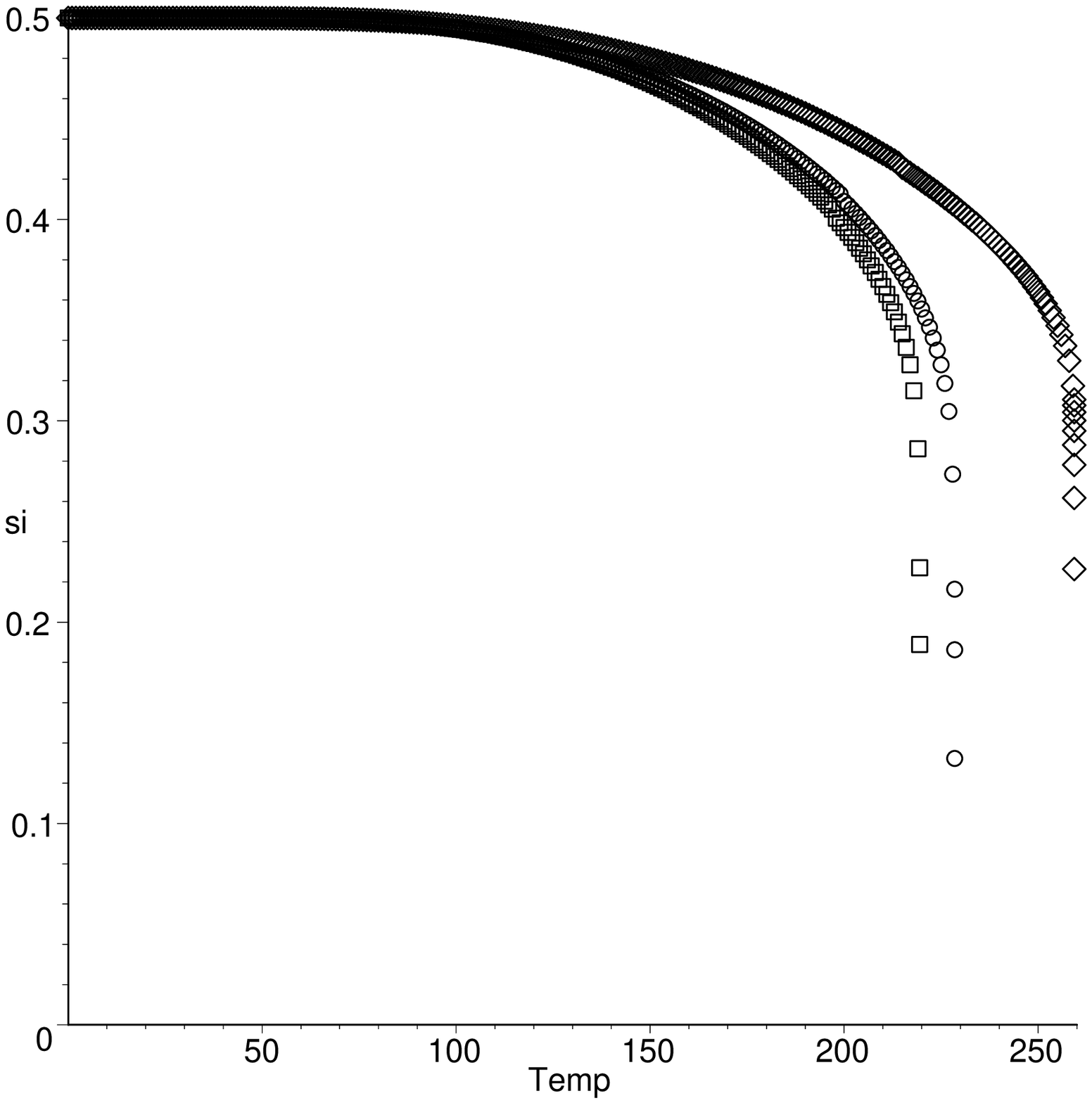}
\caption{Temperature dependence of the polarization ${\sigma}$ for $J_b = 495 K$, $\Omega_b= 20 K$,
$J_s=900 K$, $\Omega_s=\Omega_b$, $N$=7 and different defect layers: (1) $J_4=J_b=495 K,\, T_c =220 (\boxempty )$, 
(2) $J_4=J_d=1000 K,\, T_c = 229 (\circ )$, (3) $J_3=J_4=J_5=1000 K,\, T_c =259 K (\diamond )$.}
\label{Fig.6}
\end{figure}

The polarization is depending, too, on that how many inner layers are defect. This is shown 
in Figs.~\ref{Fig.5}, \ref{Fig.6} for a thin film with $N=7$ layers. In Fig.~\ref{Fig.5} is demonstrated that 
for $J_d=200K$ with increasing of the number of defect layers (curve 1 - $J_4=J_d=J_b$,
curve 2 - $J_4=J_d=200 K$, curve 3 - $J_3=J_4=J_5=J_d=200K$) the polarization and the critical 
temperature decreases. The opposite behavior can be seen in Fig.~\ref{Fig.6}, where $J_d=1000K$.
With increasing of the number of defect layers the polarization and $T_C$ increase.
The spin-wave energies are reduced in the first case and enhanced in the second case.

\begin{figure} [!ht]
\centering
\psfrag{si}[][][1.3]{${\sigma}$}
\psfrag{Temp}[][][0.75]{$T [K]$}
\includegraphics[scale=0.5]{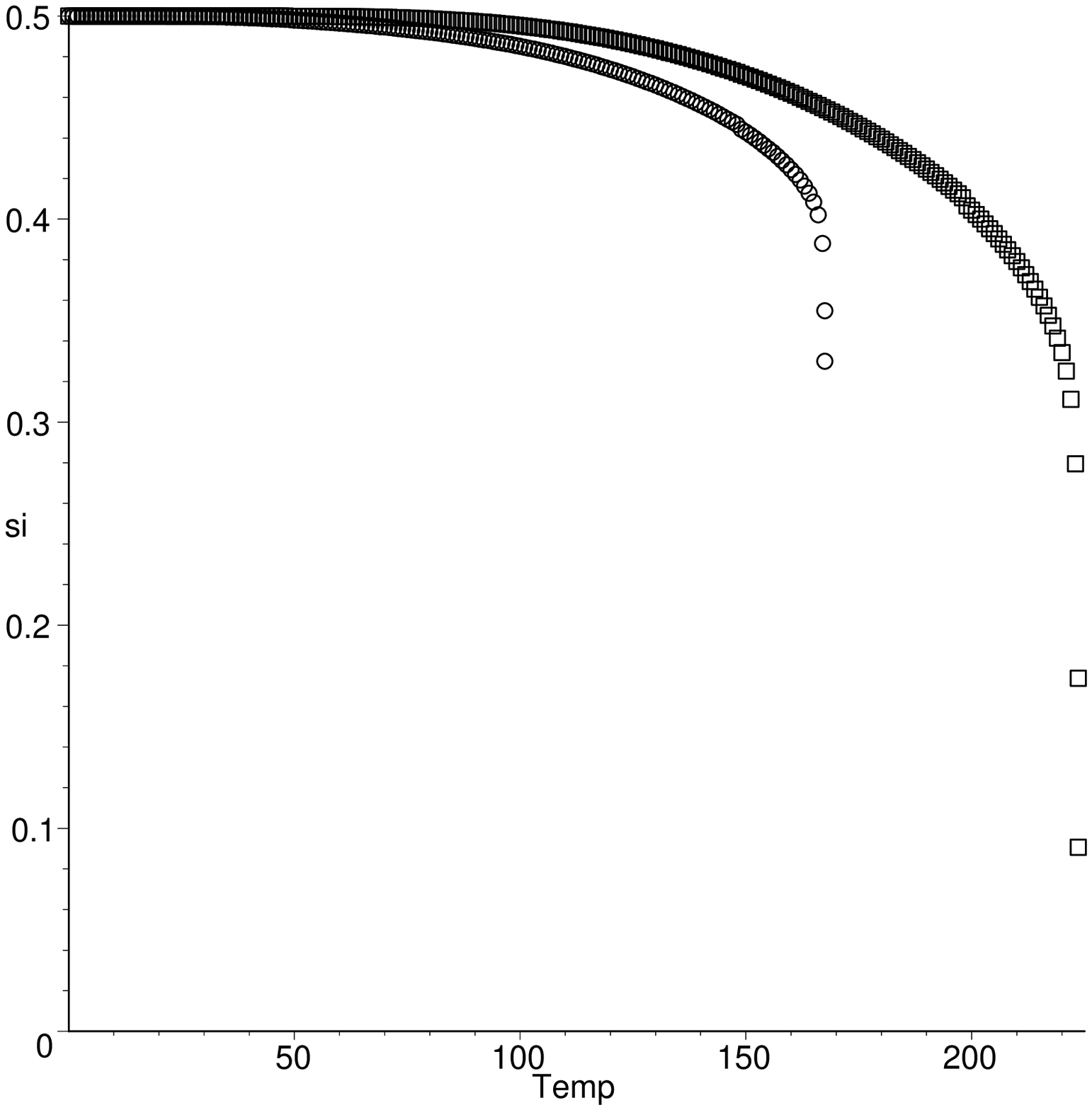}
\caption{Temperature dependence of the polarization ${\sigma}$ with $J_b = 495 K$, $\Omega_b= 20 K$,
$J_s=900 K$, $\Omega_s=\Omega_b$, $N$=9 and an antiferroelectric defect layer: (1) $J_d=1000 K,\, T_c =223 K (\boxempty )$, 
(2) $-1000 K,\, T_c =168 K (\circ )$.}
\label{Fig.7}
\end{figure}

\begin{figure} [!ht]
\centering
\psfrag{si}[][][1.3]{${\sigma}$}
\psfrag{Temp}[][][0.75]{$T [K]$}
\includegraphics[scale=0.5]{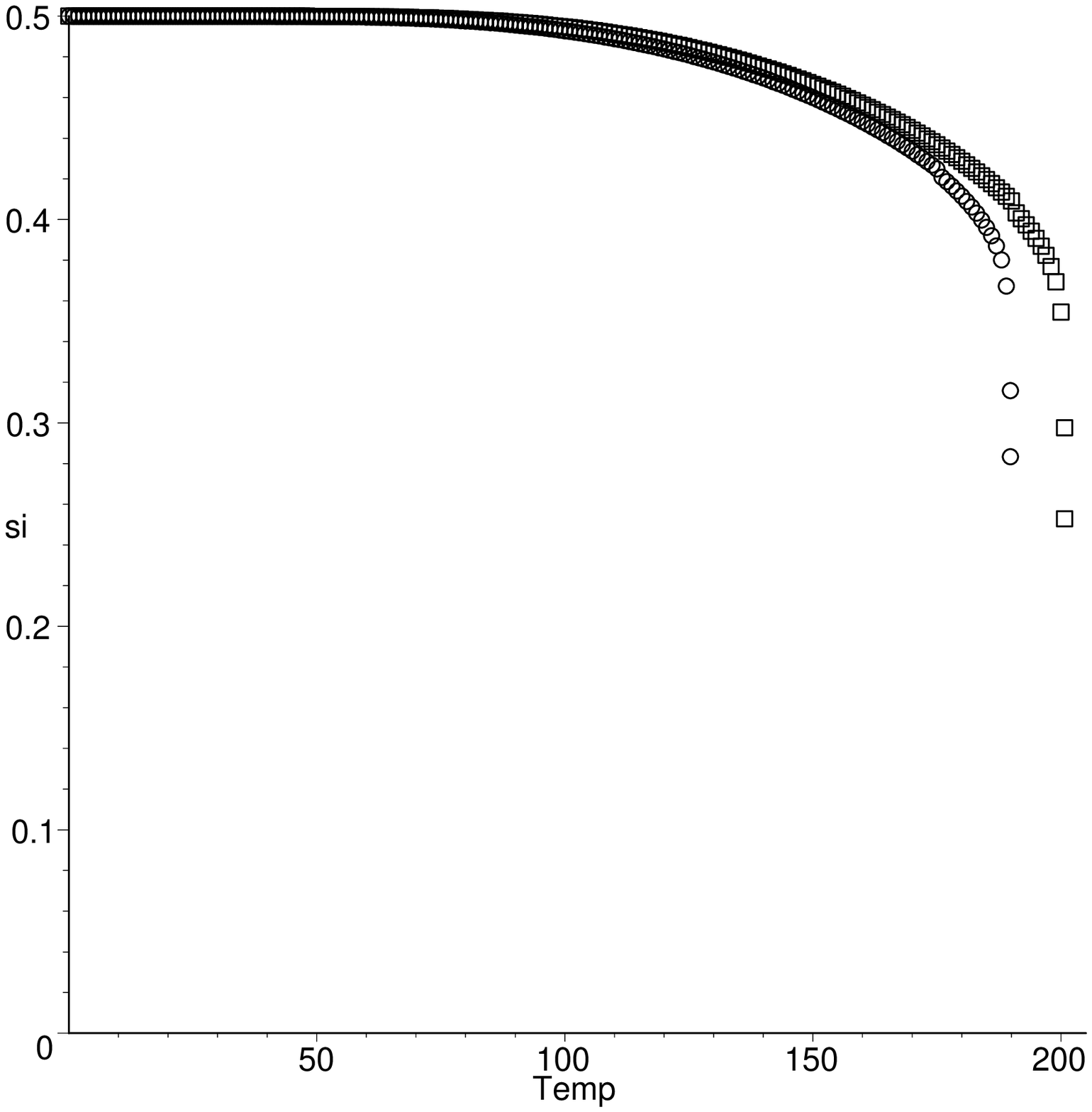}
\caption{Temperature dependence of the polarization ${\sigma}$ with $J_b = 495 K$, $\Omega_b= 20 K$,
$J_s=900 K$, $\Omega_s=\Omega_b$, $N$=9 and an antiferroelectric defect layer: (1) $J_d=300 K,\, T_c =201 K (\boxempty )$, 
(2) $-300 K,\, T_c =190 K (\circ )$.}
\label{Fig.8}
\end{figure}

We have also studied a ferroelectric thin film whit $N=9$ layers where one layer, the 
middle one is defect and the spins are ordered antiparallel, the defect layer is 
antiferroelectric, $J_5=J_d=-1000 K$ (Fig.~\ref{Fig.7}, curve 2) and $J_5=J_d=-300 K$ (Fig.~\ref{Fig.8}, curve 2). 
The polarization and $T_C$ of the film are smaller in comparison with the case where the middle 
layer is defect, but with ferroelectric order. The spin-wave energy of the thin film is also 
reduced when $J_5=J_d=-1000K$. Such a sublattice model with ferroelectric and antiferroelectric 
layers where one or more layers are defect will be studied in a next paper.

\section{Conclusions}
In conclusions, based on a modified transverse Ising model and using a Green's function technique,
the polarization, the spin-wave energies and the phase transition temperature for ferroelectric 
thin films with structural defects are calculated. The dependence on temperature, film thickness 
and interaction constants is discussed. It is shown that defect layers in FE thin films, layers 
with impurities or vacancies, with dislocations, can induce strong increasing or decreasing of 
the polarization, the spin-wave energies and of the critical temperature of FE phase transition 
due to different exchange interactions in the defect layers. The results are in good agreement 
with the experimental data for different ferroelectrics.

\begin{acknowledgments} 
One of us (J. M. W.) is grateful to the Deutsche Forschungsgemeinschaft 
for financial support. This work is supported by the SFB 418. Further, we thank Dr. D. Hesse and 
Dr. M. Alexe, Max Planck Institute of Microstructure Physics, Halle for fruitful and exciting discussions  
on experimental details.
\end{acknowledgments}

\newpage

\end{document}